\documentclass[onecolumn,prd,aps,12pt]{revtex4}
\usepackage{amsmath}
\usepackage{amsfonts}
\usepackage{amssymb}
\usepackage{graphicx}

\begin{document}

\title{Casimir Effect at finite temperature for the Kalb-Ramond field}
\author{H. Belich $^{a,b,c}$, L.M. Silva$^{b,d}$, J.A. Helay\"{e}l-Neto$%
^{c,e}$, A.E. Santana$^{b}$, }
\affiliation{$^{a}${\small {\ Departamento de F\'{\i}sica e Qu\'{\i}mica, Universidade
Federal do Esp\'{\i}rito Santo, Vit\'{o}ria, ES, 29060-900, Brazil}}}
\affiliation{$^{b}${\small {Instituto de F\'{\i}sica, International Center for Condensed
Matter Physics, Universidade de Bras\'{\i}lia, 70910-900, Bras\'{\i}lia, DF,
Brazil }}}
\affiliation{{\small \ }$^{c}${\small {Grupo de F\'{\i}sica Te\'{o}rica Jos\'{e} Leite
Lopes, C.P. 91933, 25685-970, Petr\'{o}polis, RJ, Brazil}}}
\affiliation{{\small {~}}$^{d}${\small {\ Departamento de Ci\^{e}ncias Exatas e da Terra,
Universidade do Estado da Bahia, Campus II, 48040210 , Alagoinhas, BA, Brazil%
}}}
\affiliation{$^{e}$ {\small {Centro Brasileiro de Pesquisas F\'{\i}sicas, Rua Xavier
Sigaud 150, Rio de Janeiro, RJ, 22290-180, Brazil}}}
\email{belichjr@gmail.com, lourivalmsfilho@uol.com.br, helayel@cbpf.br,
asantana@unb.br }
\date{\today}

\begin{abstract}
We use the thermofield dynamics (TFD) formalism to obtain the
energy-momentum tensor for the Kalb-Ramond (KR) field in a topology $%
S^{1}\times S^{1}\times R^{2}$. The compactification is carried out by a
generalized TFD-Bogoliubov transformation that is used to define a
renormalized energy-momentum tensor. The expressions for the Casimir energy
and pressure at finite temperature are then derived. A comparative analysis
with the electromagnetic case is developed, and the results may be important
for applications, as in cuprate superconductivity, for instance.
\end{abstract}

\maketitle

\section{Introduction}

The Casimir effect is one of the most remarkable manifestations of the
vacuum fluctuations. For the electromagnetic field, it consists in the
attraction between two metallic plates, parallel each other, embedded\ into
the vacuum \cite{Casimir48}. The attraction is due to a fluctuation of the
fundamental energy of the field caused by the presence of planes, which
select the electromagnetic vacuum modes by boundary conditions \cite%
{Milton,Miloni,Plunien,Mostepanenko}. In general, the Casimir effect is then
a modification in the vacuum energy of a given quantum field due to the
imposition of boundary conditions or topological effects on this field.

For the case of the electromagnetic field, the measurement in great accuracy
in the last decade has gained attention of the theoretical and experimental
community \cite{Lamo, Mohi}. One practical implication of these achievements
is the development of nanodispositives. For instance, recently, it has
appeared a possible implication of Casimir effect in high-$T_{c}$
superconductors: the role of the Casimir plates can be attributed to the $%
nCuO_{2}$ layers, which form a Cu-O non-superconducting charge carriers
layers initially, and are able to form the superconductors layers below the
critical temperature, $T_{c}$ \cite{kempf,bordag,mtdo1}.

In cuprate superconductivity, a vortex-boson duality takes place in the Cu-O
layers~\cite{Tesa}. This duality relates the description of 2-dimension
phase-fluctuating superconductor with an Abrikosov vortex lattice of a dual
fictitious superconductor. The duality is based on the fact that if a vortex
is fixed, the Cooper pairs propagates coherently. If a vortex moves, the
phase of condensate is more uncertain, and the coherent propagation \ of
pairs is frustrated. If this duality is extended to 3 dimensions, the vortex
becomes a string and the gauge field is described by a Kalb-Ramond (KR)
field, a rank-2 skew-symmetric tensor field \cite{KRref,Franz}. Since the
distances of the layers are of the order of few nanometers, the Casimir
effect for the electromagnetic field is remarkable and has been analyzed.
For the same motive, it is of interest to investigate the Casimir effect for
the KB field \cite{Leo}, including the temperature effect. In this paper, we
address this problem, keeping in mind that such a study may be important in
string theory and supergravity, as well, where the KB field also arises
naturally.

We\ start by reviewing some aspect of the KR field \cite{KRref}, \emph{\ }%
submitted to periodic boundary conditions, in order to compare our results
with standard situations presented in the literature \cite{Leo}. The
temperature effect is introduced with the real-time formalism in the
canonical version, i.e. thermofield dynamics (TFD)~\cite{3ume1}. In this
case, the thermal theory is constructed on a Hilbert space and temperature
is introduced by a Bogoliubov transformation~\cite{3ume2,3das,kha1}. \
Considering that a thermal field theory is a quantum field compactified in a
topology $S^{1}\times R^{3}$, a result of the KMS (Kubo, Martin, Schwinger)
condition, this apparatus has been used to describe field theories in
toroidal topologies \cite{birrel1,birrel2,3comp1,3comp2,3comp3,3comp4}. In
terms of TFD, the Bogoliubov transformation has then been generalized to
describe thermal and space-compactification effects with real (not
imaginary) time. Here we consider a Bogoliubov transformation to take into
account the KR field in a topology $S^{1}\times S^{1}\times R^{2}$. Such a
mechanism is quite suitable to treat, in particular, the Casimir effect.
This is a consequence of the nature of the propagator that is written in two
parts: one describes the flat (Minkowsky) space-time contribution, whilst
the other addresses to the thermal and the topological effect. In such a
case, a renormalized energy-momentum tensor is introduced in a consistent
and simple way \cite{kha1}. For the Casimir effect, it is convenient to work
with the real-time canonical formalism, although the 2-form gauge field is
not commonly studied in the context of canonical quantization. For this
reason, we calculate explicitly the KR energy-momentum tensor in terms of
the propagator.

The paper is organized in the following way. In Section II, some aspects of
TFD are presented to describe a field in a topology $S^{1}\times S^{1}\times
R^{3}$. In Section III, the KR energy-momentum tensor is derived. In Section
IV, the topology $S^{1}\times S^{1}\times R^{3}$ is considered; and Section
V, the Casimir effect for the KR field is studied. Concluding remarks are
presented in Section VI.

\section{Thermofield dynamics and topology $S^{1}\times S^{1}\times R^{3}$}

In this section we present some elements of thermofield dynamics(TFD),
emphasizing aspects to be used in the calculation of the Casimir effect for
the Kalb-Ramon field. In short, TFD is introduced by two basic ingredients
\cite{kha1}. Considering a von-Neumann algebra of operator in Hilbert space,
there is a doubling, corresponding to the commutants introduced by a modular
conjugation. This corresponds to a doubling of the original Fock space of
the system leading to the expanded space $\mathcal{H}_{T}=\mathcal{H}\otimes%
\widetilde{\mathcal{H}}$. This doubling is defined by a mapping $\widetilde{}%
:\mathcal{H}\rightarrow\widetilde{\mathcal{H}}$, associating each operator $%
a $ acting on $\mathcal{H}$ with two operators in $\mathcal{H}_{T}$, $A\ $
and $\tilde{A}$, which are connected by the modular conjugation in a c$%
^{\ast}$-algebra, also called tilde conjugation rules \cite{kha2,kha3}:
\begin{align*}
(A_{i}A_{j})\widetilde{} & =\widetilde{A}_{i}\widetilde{A}_{j}, \\
(cA_{i}+A_{j})\widetilde{} & =c^{\ast}\widetilde{A}_{i}+\widetilde{A}_{j}, \\
(A_{i}^{\dagger})\widetilde{} & =(\widetilde{A}_{i})^{\dagger}, \\
(\widetilde{A}_{i})^{\widetilde{}} & =-\xi A_{i},
\end{align*}
with $\xi=-1$ for bosons and $\xi=+1$ for fermions. The physical variables
are described by nontilde operators. The tilde variables, defined in the
commutant of the von Neumann algebra, are associated with generators of the
modular group given by $\widehat{A}=A-\widetilde{A}$. With this elements,
reducible representations of Lie-groups can be studied, in particular,
kinematical symmetries as the Lorentz group. This gives rise to
Liouville-von-Neumann-like equations of motion. The other basic ingredient
of TFD is a Bogoliubov transformation, $B(\alpha)$, introducing a rotation
in the tilde and non-tilde variables, such that thermal effects emerge from
a condensate state. The rotation parameter $\alpha$ is associated with
temperature, and this procedure is equivalent to the usual statistical
thermal average. \ In the standard doublet notation \cite{3ume2}, we write
\begin{equation}
(A^{r}(\alpha))=\left(
\begin{array}{c}
A(\alpha) \\
\xi\widetilde{A}^{\dagger}(\alpha)%
\end{array}
\right) =B(\alpha)\left(
\begin{array}{c}
A \\
\,\xi\widetilde{A}^{\dagger}%
\end{array}
\right) \,,
\end{equation}
$(\,A^{r}(\alpha))^{\dagger}=\left( A^{\dagger}(\alpha)\,,\,\widetilde {A}%
(\alpha)\right) \,$, with the Bogoliubov transformation given by
\begin{equation}
B(\alpha)=\left(
\begin{array}{cc}
u(\alpha) & -v(\alpha) \\
\xi v(\alpha) & \,\,\,\,u(\alpha)%
\end{array}
\right) ,  \label{BT}
\end{equation}
where $\,u^{2}(\alpha)+\xi v^{2}(\alpha)=1$.

The parametrization of the Bogoliubov transformation in TFD is obtained by
setting $\alpha=\beta=T^{-1}$ and by requiring that the thermal average of
the number operator, $N(\alpha)=a^{\dagger}(\alpha)a(\alpha)$, i.e. $\langle
N(\alpha)\rangle_{\alpha}=\langle0,\tilde{0}|a^{\dagger}(\alpha)a(\alpha )|0,%
\tilde{0}\rangle$, \ gives either the Bose or the Fermi distribution, i.e
\begin{equation}
N(\alpha)=\,\,\,v^{2}(\beta)=\left( e^{\beta\varepsilon}+\xi\right) ^{-1}\,.
\label{UV}
\end{equation}
Here we have used, for the sake of simplicity of notation, $A\equiv a$ and $%
\widetilde{A}\equiv\widetilde{a}$, and
\begin{equation*}
a=u(\alpha)a(\alpha)+v(\alpha)\,\widetilde{a}^{\dagger}(k,\alpha),
\end{equation*}
such that the other operators ($a^{\dag}(k),\widetilde{a}(k),\widetilde {a}%
^{\dagger}(k)$) can be obtained by applying the hermitian or the tilde
conjugation rules. It is shown then that the thermal average, $\langle
N(\alpha)\rangle_{\alpha}$, can be written as $\langle N(\alpha)\rangle
_{\alpha}=\langle0(\alpha)|a^{\dagger}a|0(\alpha)\rangle$, where $%
|0(\alpha)\rangle$ is given by $|0(\alpha)\rangle=U(\alpha)|0,\widetilde {0}%
\rangle$, with
\begin{equation*}
U(\alpha)=\exp\{\theta(\alpha)[a^{\dag}\widetilde{a}^{\dag}-a\widetilde{a}%
]\}.
\end{equation*}

Let us consider the free Klein-Gordon field described by the Hamiltonian $%
\mathcal{H}=\frac{1}{2}\partial_{\alpha}\phi\partial^{\alpha}\phi-\frac{1}{2}%
m^{2}\phi^{2},$ in a Minkowski space specified by the diagonal metric with
signature $(+,---)$. The generalization of $U(\alpha)$ is then defined for
all modes, such that%
\begin{align*}
\phi(x;\alpha) & =U(\alpha)\phi(x)U^{-1}(\alpha), \\
\widetilde{\phi}(x;\alpha) & =U(\alpha)\widetilde{\phi}(x)U^{-1}(\alpha).
\end{align*}
Using a Bogoliubov transformation for each mode, we get \cite{kha1}
\begin{equation*}
\phi(x;\alpha)=\int\frac{d^{3}k}{(2\pi)^{3}}\frac{1}{2k_{0}}\ [a(k;\alpha
)e^{-ikx}+a^{\dag}(k;\alpha)e^{ikx}]
\end{equation*}
and%
\begin{equation*}
\widetilde{\phi}(x;\alpha)=\int\frac{d^{3}k}{(2\pi)^{3}}\frac{1}{2k_{0}}\ [%
\widetilde{a}(k;\alpha)e^{ikx}+\widetilde{a}^{\dag}(k;\alpha)e^{-ikx}].
\end{equation*}

The $\alpha$-propagator is defined by
\begin{align}
G(x-y,\alpha) & =-i\langle0,\widetilde{0}|\mathrm{T}[\phi(x;\alpha
)\phi(y;\alpha)]|0,\widetilde{0}\rangle  \notag \\
& =-i\langle0(\alpha)|\mathrm{T}[\phi(x)\phi(y)]|0(\alpha)\rangle,
\label{T3}
\end{align}
where T is the time-ordering operator. This leads to%
\begin{equation}
G_{0}(x-y,\alpha)=\int\frac{d^{4}k}{(2\pi)^{4}}e^{-ik(x-y)}\ G_{0}(k,\alpha),
\label{T2}
\end{equation}
where%
\begin{equation}
G_{0}(k;\alpha)=G_{0}(k)+\ v^{2}(k_{\alpha};\alpha)[G_{0}(k)-G_{0}^{\ast
}(k)],  \label{T1}
\end{equation}
with
\begin{equation*}
G_{0}(k)=\frac{1}{k^{2}-m^{2}+i\varepsilon},
\end{equation*}
such that%
\begin{equation*}
G_{0}(k)-G_{0}^{\ast}(k)=2\pi i\delta(k^{2}-m^{2}).
\end{equation*}
Using $v^{2}(k_{\alpha};\alpha)=v^{2}(k^{0};\beta)$ as the boson
distribution, $n(k^{0};\beta)$, i.e.
\begin{equation}
v^{2}(k^{0};\beta)=n(k^{0};\beta)=\frac{1}{(e^{\beta\omega_{k}}-1)}%
=\sum\limits_{l_{0}=1}^{\infty}e^{-\beta k^{0}l_{0}},  \label{june121}
\end{equation}
with $\omega_{k}=k_{0}$ and $\beta=1/T$, $T$ being the temperature, then we
have
\begin{equation}
G(k,\beta)=G_{0}(k)+2\pi i~n(k^{0},\beta)\delta(k^{2}-m^{2}),  \label{mats71}
\end{equation}
with%
\begin{equation*}
G_{0}(x-y)=\int\frac{d^{4}k}{(2\pi)^{4}}e^{-ik(x-y)}\ G_{0}(k).
\end{equation*}
For the case $m=0$, we have
\begin{equation}
G_{0}(x-y)=\frac{-i}{(2\pi)^{2}}\frac{1}{(x-x^{\prime})^{2}-i\varepsilon },
\label{june1}
\end{equation}
\

The Green function given in Eq. (\ref{T3}) is also written as
\begin{align*}
G_{0}(x-y,\beta) & =\mathrm{Tr}[\rho(\beta)\mathrm{T}[\phi(x)\phi(y)]] \\
& =G_{0}(x-y-i\beta n_{0},\beta),
\end{align*}
where $\rho(\beta)$ is the equilibrium density matrix for the
grand-canonical ensemble and $n_{0}=(1,0,0,0)$. This shows that $%
G_{0}(x-y,\beta)$ is a periodic function in the imaginary time, with period
of $\beta$; and the quantities $w_{n}=2\pi n/\beta$ are the Matsubara
frequencies. This periodicity is known as the KMS (Kubo, Martin, Schwinger)
boundary condition. From Eq. (\ref{T2}) we show that $G_{0}(x-y,\beta)$ is a
solution of the Klein-Gordon equation: with $\tau=it,$ such that $%
\square+m^{2}=-\partial_{\tau}^{2}-\nabla^{2}+m^{2}$, and
\begin{equation}
(\square+m^{2})G_{0}(x,\beta)=-\delta(x).  \label{T7}
\end{equation}
Then $G_{0}(x-y,\beta)$ can also be written as a in a Fourier series, i.e.
\begin{equation}
G_{0}(x-y,\beta)=\frac{-1}{i\beta}\sum_{n}\int d^{3}p\frac{e^{-ik_{n}\cdot x}%
}{k_{n}^{2}-m^{2}+i\varepsilon},  \label{8propmat1}
\end{equation}
where $k_{n}=(k_{n}^{0},\mathbf{k}).$ The propagator, given in Eq.~(\ref{T3}%
) and in Eq.~(\ref{8propmat1}), is solution of Eq.~(\ref{T7}) and fullfils
the same boundary condition of periodicity and Feyman contour. Then these
solutions are the same. A direct proof is provided by Dolan and Jackiw in
the case of temperature~\cite{DJ1}.

Due to the periodicity and the fact $G_{0}(x,\beta)$ and $G_{0}(x-y)$
satisfy Eq.~(\ref{T7}), the same local structure, then this finite
temperature theory results to be the $T=0$ theory compactified in a topology
$\Gamma_{4}^{1}=S^{1}\times\mathbb{R}^{3}$, where the (imaginary) time is
compactified in $S^{1}$, with circumference $\beta$. The Bogoliubov
transformation introduces the imaginary compactification through a
condensate.

For an Euclidian theory, this procedure can be developed for space
compactification. Considering a compactification along the axis $x^{1}$, we
have \cite{kha1}
\begin{equation}
G_{0}(x-y,L_{1})=\frac{1}{L_{1}}\sum_{l=-\infty}^{\infty}\int d^{3}k\ \frac{%
-e^{-ik(x-y)}}{k^{2}-m^{2}+i\varepsilon}.  \label{bo6}
\end{equation}
This Green function is written as $G_{0}(x-y,\alpha)$ in Eq. (\ref{T2}), i.e.%
\begin{align}
G_{0}(x-y,L_{1}) & =\int\frac{d^{4}k}{(2\pi)^{4}}e^{-ik(x-y)}  \notag \\
& \times\{G_{0}(k)+v^{2}(k^{1};L_{1})[G_{0}(k)-G_{0}^{\ast}(k)]\},
\label{T9}
\end{align}
with the Bogoliubov transformation given by%
\begin{equation}
v^{2}(k^{1},L_{1})=\sum\limits_{n=1}^{\infty}e^{-inL_{1}k^{1}}.
\label{june125}
\end{equation}

From this result, we compactify this theory in the imaginary time in order
to take into account the temperature effect. Starting from Eq. (\ref{T9}),
we consider now the topology $\Gamma_{4}^{2}=S^{1}\times S^{1}\times R^{2}$.
The boson field is compactified in two directions, i.e. $x^{0}$ and $x^{1}$.
In the $x^{1}$-axis, the compactification is in a circle of circumference $%
L_{1}$ and in the Euclidian $x^{0}$- axis, the compactification is in a
circumference $\beta$, such that in both of the cases the Green function
satisfies periodic boundary conditions. In this case, we have $\ $
\begin{align}
G_{0}(x-y;\beta,L_{1}) & =\frac{1}{L_{1}}\sum\limits_{l=-\infty}^{\infty }%
\frac{1}{\beta}\sum_{n=-\infty}^{\infty}\ \frac{1}{(2\pi)^{2}}  \notag \\
& \times\int dk_{2}dk_{3}e^{-ik_{nl}(x-y)}G_{0}(k_{nl};L_{1}),
\label{t12211}
\end{align}
where%
\begin{equation*}
k_{nl}=(k_{n}^{0},k_{l}^{1},k^{2},k^{3}),\ \
\end{equation*}
with
\begin{equation*}
\ k_{n}^{0}=\frac{2\pi n}{\beta};\ \ k_{l}^{1}=\frac{2\pi l}{L_{1}}
\end{equation*}
and
\begin{equation*}
G_{0}(k_{nl};\beta,L_{1})=\frac{-1}{k_{nl}^{2}-m^{2}}.
\end{equation*}

The Green function in the form of a Fourier integral is
\begin{align*}
G(x-y;\beta,L_{1}) & =\int\frac{d^{4}k}{(2\pi)^{4}}e^{-ik(x-y)}\{G_{0}(k) \\
& +v^{2}(k_{0},k_{1};\beta,L_{1})[G_{0}(k)-G_{0}^{\ast}(k)]\},
\end{align*}
where%
\begin{align}
v^{2}(k^{0},k^{1};\beta,L_{1}) & =v^{2}(k^{0};\beta)+v^{2}(k^{1};L_{1})
\notag \\
& +2v^{2}(k^{0};\beta)v^{2}(k^{1};L_{1}).  \label{june124}
\end{align}
This corresponds to a generalization of the Dolan-Jackiw propagator,
describing a system of free bosons at finite temperature, with a
compactified space dimension~\cite{kha1,DJ1}. Observe the following
consistency relations
\begin{align*}
v_{B}^{2}(k^{0};\beta) &
=\lim_{L_{1}\rightarrow\infty}v^{2}(k^{0},k^{1};\beta,L_{1}), \\
v_{B}^{2}(k^{1};L_{1}) &
=\lim_{\beta\rightarrow\infty}v^{2}(k^{0},k^{1};\beta,L_{1}).
\end{align*}
In the next sections we use these results to analyze the energy-momentum
tensor of the Kalb-Ramon field.

\section{Kalb-Ramon energy-momentum tensor}

The Kalb-Ramon Lagrangian density is given by\emph{\ }%
\begin{equation}
\mathcal{L}=\frac{1}{3!}G_{\mu\nu\gamma}G^{\mu\nu\gamma}  \label{lagranKR}
\end{equation}
where,\emph{\ }%
\begin{equation}
G_{\alpha\mu\nu}=\partial_{\alpha}B_{\mu\nu}+\partial_{\mu}B_{\nu\alpha
}+\partial_{\nu}B_{\alpha\mu,}
\end{equation}
is the field strength for the KR field, $B^{\mu\nu}$. It is worthy
emphasizing that $B^{\mu\nu}=-B^{\nu\mu}$, and that the Lagrangian density
given in Eq. (\ref{lagranKR}) is invariant by the gauge transformation
\begin{equation}
B^{\mu\nu}(x)\rightarrow B{^{\prime}}^{\mu\nu}(x)=B^{\mu\nu}(x)+\partial^{%
\mu }\Lambda^{\nu}(x)-\partial^{\nu}\Lambda^{\mu}(x)\ ,  \label{zxc1}
\end{equation}
where $\Lambda^{\mu}$ is an arbitrary vector field. The operator
energy-momentum tensor is given by

\begin{equation}
T^{\mu\nu}=\frac{3}{2}G^{\mu\alpha\beta}G_{\text{ \ }\alpha\beta}^{\nu }%
\mathcal{-}\frac{1}{4}g^{\mu\nu}G^{\alpha\beta\gamma}G_{\alpha\beta\gamma}.
\end{equation}

The canonical conjugated moment related to $B_{\mu\nu}$ is

\begin{equation}
\pi_{\nu\kappa}=G_{0\nu\kappa}.
\end{equation}
Adopting the Lorentz gauge condition $\partial_{\mu}B^{\nu\kappa}=0$, and
adding in the Lagrangian density the \textquotedblleft Fermi
term\textquotedblright

\begin{equation}
\frac{1}{4}\left( \partial _{\nu }B^{\nu \kappa }\partial ^{\mu }B_{\mu
\kappa }+\partial _{\nu }B^{\kappa \nu }\partial ^{\mu }B_{\kappa \mu
}\right) =0,
\end{equation}%
we can carry out a covariant quantization. Then the commutation relations
become
\begin{align*}
\left[ \pi _{0i},B^{0j}\right] & =\left[ \pi _{0i},B^{jk}\right] =\left[ \pi
_{ij},B^{0k}\right] =0, \\
\left[ B_{\mu \nu },B^{\rho \sigma }\right] & =\left[ \pi _{\mu \nu },\pi
^{\rho \sigma }\right] =0, \\
\left[ \pi _{ij}(\mathbf{x},t),B^{kl}(\mathbf{x}{\acute{}},t)\right] &
=-i\delta _{ij}^{kl}\left( \mathbf{x},\mathbf{x}{\acute{}}\right) ,\text{ }%
i,j,k=1,2,3,
\end{align*}%
where%
\begin{equation*}
\text{\ }\delta ^{ijkl}(\mathbf{x}-\mathbf{x}{\acute{}})=\int \frac{d^{3}q}{%
\left( 2\pi \right) ^{3}}\exp i\mathbf{q}\left( \mathbf{x}-\mathbf{x}{\acute{%
}}\right) \left[ N^{ijkl}-H^{ijkl}(q)\right] \text{\ ,}
\end{equation*}

\begin{equation}
N^{ijkl}=\frac{1}{2}\left( \delta ^{ik}\delta ^{jl}-\delta ^{il}\delta
^{jk}\right) \
\end{equation}%
and

\begin{equation}
H^{ijkl}\left( q\right) =\frac{1}{2q^{2}}\left[ \left( q^{i}q^{k}\delta
^{jl}-q^{i}q^{l}\delta ^{jk}\right) -\left( q^{j}q^{k}\delta
^{il}-q^{j}q^{l}\delta ^{ik}\right) \right] ,\text{ \ \ }q^{2}=\mathbf{q}%
\cdot \mathbf{q}.
\end{equation}%
Beyond that, we have

\begin{align*}
\partial _{i}\pi ^{ij}& =\left( \nabla \times \mathbf{N}\right) ^{j}=0;\text{
\ \ }N_{i}=\tilde{G}_{i}\text{ \ \ }K_{0}=\tilde{G}_{0}, \\
\partial _{i}\delta ^{ijk}(\mathbf{x}-\mathbf{x}{\acute{}})& =0,\text{ \ }%
\partial {\acute{}}_{k}\delta ^{ijkl}(\mathbf{x}-\mathbf{x}{\acute{}})=0.
\end{align*}%
After some calculations, one reaches

\begin{equation}
\left[ \pi _{ij}(\mathbf{x},t),\phi ^{kl}\left( \mathbf{x}{\acute{}}%
,t\right) \right] =-i\delta _{ij}^{kl}\left( \mathbf{x},\mathbf{x}{\acute{}}%
\right) ,\text{ \ \ }\left[ \phi _{ij}(\mathbf{x},t),\pi ^{kl}\left( \mathbf{%
x}{\acute{}},t\right) \right] =i\delta _{ij}^{kl}\left( \mathbf{x},\mathbf{x}%
{\acute{}}\right) .
\end{equation}%
The operator energy-momentum tensor for the KR field is
\begin{equation}
T^{\mu \nu }=\frac{3}{2}G^{\mu \alpha \beta }G_{\text{ \ }\alpha \beta
}^{\nu }\mathcal{-}\frac{1}{4}g^{\mu \nu }G^{\alpha \beta \gamma }G_{\alpha
\beta \gamma }
\end{equation}

The operator $T^{\mu\nu}$ is written as%
\begin{align*}
T^{\mu\nu}(x) & =\lim_{x\rightarrow x%
{\acute{}}%
}T\left[ \frac{3}{2}G^{\mu\alpha\beta}(x)G_{\text{ \ }\alpha\beta}^{\nu}(x%
{\acute{}}%
)\mathcal{-}\frac{1}{4}g^{\mu\nu}G^{\alpha\beta\gamma}(x)G_{\alpha\beta%
\gamma }(x%
{\acute{}}%
)\right] \\
& =\lim_{x\rightarrow x%
{\acute{}}%
}\left[ \frac{3}{2}\mathcal{G}^{\mu\alpha\beta,\nu}{}_{\alpha\beta}(x,x%
{\acute{}}%
)\mathcal{-}\frac{1}{4}g^{\mu\nu}\mathcal{G}^{\alpha\beta\gamma}{}_{\alpha%
\beta\gamma}(x,x%
{\acute{}}%
)\right] ,
\end{align*}
where%
\begin{equation}
\mathcal{G}^{\alpha\beta\gamma,\mu\nu\rho}{}(x,x%
{\acute{}}%
)=\mathrm{T}[G^{\alpha\beta\gamma}(x)G^{\mu\nu\rho}(x%
{\acute{}}%
)].
\end{equation}
Using the $\mathrm{T}$ operator explicitly, we write

\begin{equation}
\mathcal{G}^{\alpha\beta\gamma,\mu\nu\rho}{}(x,x%
{\acute{}}%
)=G^{\alpha\beta\gamma}(x)G_{\text{ \ }}^{\mu\nu\rho}(x%
{\acute{}}%
)\theta(x^{0}-x%
{\acute{}}%
^{0})\mathcal{+}G_{\text{ \ }}^{\mu\nu\rho}(x%
{\acute{}}%
)G^{\alpha\beta\gamma}(x)\theta(x%
{\acute{}}%
^{0}-x^{0}).
\end{equation}

Performing a long, but straightforward, calculation, we obtain%
\begin{align*}
\mathcal{G}^{\alpha \beta \gamma ,\mu \nu \rho }{}(x,x%
{\acute{}}%
)& =\Xi ^{\alpha \beta \gamma ,\mu \nu \rho }(x,x^{\prime })-\eta
_{0}^{\alpha }\delta (x^{0}-x%
{\acute{}}%
^{0})I^{\beta \gamma ,\mu \nu \rho }(x,x%
{\acute{}}%
) \\
& +\eta _{0}^{\mu }\delta (x^{0}-x%
{\acute{}}%
^{0})I^{\nu \rho ,\alpha \beta \gamma }(x%
{\acute{}}%
,x)+ \\
& -\eta _{0}^{\beta }\delta (x^{0}-x%
{\acute{}}%
^{0})I^{\gamma \alpha ,\mu \nu \rho }(x,x%
{\acute{}}%
)+\eta _{0}^{\nu }\delta (x^{0}-x%
{\acute{}}%
^{0})I^{\rho \mu ,\beta \gamma \alpha }(x%
{\acute{}}%
,x)+ \\
& -\eta _{0}^{\gamma }\delta (x^{0}-x%
{\acute{}}%
^{0})I^{\alpha \beta ,\mu \nu \rho }(x,x%
{\acute{}}%
)+\eta _{0}^{\rho }\delta (x^{0}-x%
{\acute{}}%
^{0})I^{\mu \nu ,\gamma \alpha \beta }(x%
{\acute{}}%
,x)
\end{align*}%
where

\begin{equation}
\Xi^{\alpha\beta\gamma,\mu\nu\rho}(x,x^{\prime})=\Gamma^{\alpha\beta\gamma
,\mu\nu\rho,\lambda\kappa\theta\psi}(x,x^{\prime})\mathrm{T}%
[B_{\lambda\kappa }(x)B_{\theta\psi}(x%
{\acute{}}%
)],
\end{equation}
with

\begin{align*}
\Gamma ^{\alpha \beta \gamma ,\mu \nu \rho ,\lambda \kappa \theta \psi
}(x,x^{\prime })& =(g^{\beta \lambda }g^{\gamma \kappa }\partial ^{\alpha
}+g^{\gamma \lambda }g^{\alpha \kappa }\partial ^{\beta }+g^{\alpha \lambda
}g^{\beta \kappa }\partial ^{\gamma }) \\
& \times (g^{\nu \theta }g^{\rho \psi }\partial
{\acute{}}%
^{\mu }+g^{\rho \theta }g^{\mu \psi }\partial
{\acute{}}%
^{\nu }+g^{\mu \theta }g^{\nu \psi }\partial
{\acute{}}%
^{\rho }),
\end{align*}%
and%
\begin{equation}
I^{\beta \gamma ,\mu \nu \rho }(x,x%
{\acute{}}%
)=\left[ B^{\beta \gamma }(x),G_{\text{ \ }}^{\mu \nu \rho }(x%
{\acute{}}%
)\right] .
\end{equation}

The average of \ the operator in the vacuum is given by%
\begin{equation}
\left\langle T^{\mu\nu}(x)\right\rangle _{0}=\left\langle 0\left\vert
T^{\mu\nu}(x)\right\vert 0\right\rangle =-i\lim_{x\rightarrow x%
{\acute{}}%
}\Gamma^{\mu\nu}(x,x^{\prime})G_{0}(x-x^{\prime}),
\end{equation}
where
\begin{equation*}
\Gamma^{\mu\nu}(x,x^{\prime})=\frac{3}{2}(\partial^{\mu}\partial%
{\acute{}}%
^{\nu}+\frac{11}{4}g^{\mu\nu}\partial_{\alpha}\partial%
{\acute{}}%
^{\alpha})
\end{equation*}
and $G_{0}(x-x^{\prime})$ is given in Eq. (\ref{june1}).

\section{Kalb-Ramon $\protect\alpha$-energy-momentum tensor}

In this section we calculate energy-momentum for the KR field in a
compactified in a toroidal topology. We define the physical (renormalized)
energy-momentum tensor by%
\begin{equation}
\mathcal{T}^{\mu\nu}(x;\alpha)=\left\langle
T^{\mu\nu}(x;\alpha)\right\rangle _{0}-\langle T^{\mu\nu}(x)\rangle_{0}
\label{june123}
\end{equation}
where $\left\langle T^{\mu\nu}(x;\alpha)\right\rangle
_{0}=\langle0|T^{\mu\nu
}(x,\alpha)|0\rangle\equiv\langle\alpha|T^{\mu\nu}(x)|\alpha\rangle$. This
leads to
\begin{equation*}
\mathcal{T}^{\mu\nu}(x;\alpha)=-i\lim_{x\rightarrow x%
{\acute{}}%
}\Gamma^{\mu\nu}(x,x^{\prime})G_{0}(x-x^{\prime};\alpha),
\end{equation*}
where%
\begin{align*}
\overline{G}(x-x^{\prime};\alpha) &
=G_{0}(x-x^{\prime};\alpha)-G_{0}(x-x^{\prime}) \\
& =\int\frac{d^{4}k}{(2\pi)^{4}}e^{-ik(x-x^{\prime})}v^{2}(k_{\alpha},%
\alpha)[G_{0}(k)-G_{0}^{\ast}(k)].
\end{align*}

Let us calculate, as an example, the case of temperature defined by $%
\alpha=(\beta,0,0,0)$, with $v^{2}(k_{0};\beta)$ given by Eq.~(\ref{june121}%
). Then we have%
\begin{equation*}
\mathcal{T}^{\mu\nu}(\beta)=\frac{3}{8\pi^{2}}\sum_{l_{0}=1}^{\infty}\frac {1%
}{\beta^{4}l_{0}^{4}}(16n_{0}^{\mu}n_{0}^{\nu}+29g^{\mu\nu}),
\end{equation*}
where $n_{0}^{\mu}=(1,0,0,0)$. Using the Riemann Zeta function
\begin{equation*}
\zeta(4)=\sum_{l=1}^{\infty}\frac{1}{l^{4}}=\frac{\pi^{4}}{90},
\end{equation*}
we obtain%
\begin{equation*}
\mathcal{T}^{\mu\nu}(\beta)=\frac{\pi^{2}}{240\beta^{4}}(16n_{0}^{%
\mu}n_{0}^{\nu}+29g^{\mu\nu}).
\end{equation*}
This leads to the Stephan-Boltzmann law for the KR field, since the energy
and pressure are given respectively by%
\begin{equation*}
E(T)=\mathcal{T}^{00}(\beta)=\frac{3\pi^{2}}{16}T^{4};\ \ P(T)=\mathcal{T}%
^{33}(\beta)=-\frac{29\pi^{2}}{240}T^{4}.\
\end{equation*}
In the next section, we use a similar procedure to calculate the Casimir
effect.

\section{Casimir effect for the KR field}

Initially we consider the Casimir effect at zero temperature. This is given
by the KR-energy-momentum tensor $\mathcal{T}^{\mu \nu }(x;\alpha )$ given
in Eq.~(\ref{june123}), where $\alpha $ accounts for spatial
compactifications. We take $\alpha =(0,0,0,iL)$, with $L$ being the
circumference of $S^{1}$. The Bogoliubov transformation is given in Eq.~(\ref%
{june125}), that in the present notation reads
\begin{equation*}
v^{2}(k_{3};L)=\sum_{l_{3}=1}^{\infty }e^{-iLk^{3}l_{3}}.
\end{equation*}%
Thus $\mathcal{T}^{\mu \nu }(x;L)$ is given by%
\begin{equation*}
\mathcal{T}^{\mu \nu }(L)=\frac{\pi ^{2}}{240L^{4}}(16n_{3}^{\mu }n_{3}^{\nu
}+29g^{\mu \nu }),
\end{equation*}%
where $n_{3}^{\nu }=(0,0,0,1)$.

For the electromagnetic field, the Casimir effect is calculated for plates
apart from each other by a distance $a$, that is related to \thinspace$L$ by
$L=2a.$\cite{kha1} We consider this fact for the sake of comparasion. The
Casimir energy and pressure, respectively, are then given by%
\begin{align*}
E(T) & =\mathcal{T}^{00}(\beta)=\frac{29\pi^{2}}{240}\frac{1}{L^{4}}=\frac{%
29\pi^{2}}{3840}\frac{1}{a^{4}},\ \  \\
P(T) & =\mathcal{T}^{33}(\beta)=-\frac{13\pi^{2}}{240}\frac{1}{L^{4}}=-\frac{%
13\pi^{2}}{3840}\frac{1}{a^{4}}.\
\end{align*}
It is interesting to compare such a result with the Casimir effect for the
electromagnetic field. In this case, the Casimir energy and momentum are,
respectively, $E(T)=-\pi^{2}/720a^{4}$ and $P(T)=-\pi^{2}/240a^{4}$.
Therefore the Casimir effect for these two fields, the electromagnetic and
KR fields, are of the same order of magnitude. A important conclusion is
that both of fields have to be taken into account in\ the analysis of the
Casimir effect for cuprate superconductors.

The effect of temperature is introduced by taken $\alpha=(i\beta,0,0,iL)$.
Using Eq. (\ref{june124}), $v^{2}(k^{0},k^{3};\beta,L)\ $\ is given by%
\begin{align*}
v^{2}(k^{0},k^{3};\beta,L) &
=v^{2}(k^{0};\beta)+v^{2}(k^{3};L)+2v^{2}(k^{0};\beta)v^{2}(k^{3};L) \\
& =\sum\limits_{l_{0}=1}^{\infty}e^{-\beta
k^{0}l_{0}}+\sum_{l_{3}=1}^{\infty
}e^{-iLk^{3}l_{3}}+2\sum\limits_{l_{0},l_{3}=1}^{\infty}e^{-\beta
k^{0}l_{0}-iLk^{3}l_{3}}.
\end{align*}
Then the energy-momentu tensor is
\begin{align*}
\mathcal{T}^{\mu\nu}(\beta,L) & =\frac{3}{8\pi^{2}}\sum_{l_{0}=1}^{\infty }%
\frac{1}{\beta^{4}l_{0}^{4}}(16n_{0}^{\mu}n_{0}^{\nu}+29g^{\mu\nu}) \\
& +\frac{\pi^{2}}{240L^{4}}(16n_{3}^{\mu}n_{3}^{\nu}+29g^{\mu\nu}) \\
& +\frac{87}{4\pi^{2}}\sum\limits_{l_{0},l_{3}=1}^{\infty}\frac{1}{%
[(l_{3}L)^{2}-(l_{0}\beta)^{2}]^{3}} \\
& \times\{(l_{3}L)^{2}[g^{\mu\nu}+\frac{16}{29}n_{3}^{\mu}n_{3}^{\nu}] \\
& -(l_{0}\beta)^{2}[g^{\mu\nu}+\frac{16}{29}n_{0}^{\mu}n_{0}^{\nu}]\}.
\end{align*}
The Casimir energy and pressure are given, respectively, by%
\begin{align*}
E(\beta,L) & =\mathcal{T}^{00}(\beta,L)=\frac{3\pi^{2}}{16\beta^{4}}+\frac{%
29\pi^{2}}{240}\frac{1}{L^{4}} \\
& -\frac{87}{116\pi^{2}}\sum\limits_{l_{0},l_{3}=1}^{\infty}\frac{45(\beta
l_{0})^{2}-29(l_{3}L)^{2}}{[(l_{3}L)^{2}-(l_{0}\beta)^{2}]^{3}}
\end{align*}
and
\begin{align*}
P(\beta,L) & =\mathcal{T}^{33}(\beta,L)=-\frac{29\pi^{2}}{240\beta^{4}}-%
\frac{13\pi^{2}}{240}\frac{1}{L^{4}} \\
& +\frac{87}{116\pi^{2}}\sum\limits_{l_{0},l_{3}=1}^{\infty}\frac{29(\beta
l_{0})^{2}-13(l_{3}L)^{2}}{[(l_{3}L)^{2}-(l_{0}\beta)^{2}]^{3}}.
\end{align*}
The first two terms of these expressions are, respectively, the
Stephan-Boltzmann term and the Casimir effect at $T=0$. The last term
accounts for the simultaneous effect of spatial compactification, described
by $L$, and temperature, $T=1/\beta$.

\section{Concluding remarks}

In this paper we have used thermofield dynamis (TFD), a real-time formalism
\ for the thermal quantum field theory, to calculated the Casimir effect for
the Kalb-Ramon (KR) field at finite temperature. We have first derived the
expression for the energy-momentum tensor in terms of the TFD\ propagator
considering a topology $S^{1}\times S^{1}\times R^{2}$. This is used to
study the KR field compactified in spatial coordinate and at finite
temperature. In this development, the TFD apparatus is particularly
appropriated in the definition of the physical (renormalized)
energy-momentum tensor. Then the Casimir energy and pressure are calculated.

An interesting result is that the Casimir effect associated with the
kalb-Ramon field is of the same order of magnitude that the Casimir effect
for the electromagnetic field. This result is significantly useful for the
analysis of the Casimir effect in cuprate superconductors, where a
vortex-boson duality takes place in the Cu-O layers, that are apart form one
another by few nanometers. The results derived here, in addition, find
interest in string theory and supergravity, where the KR field arises
naturally.

As a gauge field, the Kalb-Ramon 2-form potential has remarkable properties
in different space-time dimensions. In (1+2)-dimensions, for instance, for
the free case, the KR field does not propagate any on-shell degree of
freedom. In (1+3)-dimensions, it may be coupled to a vector gauge potential
by means of a Chern-Simons-like term and yelds the description of a massive
spin-1 particle without any symmetry breakdown. If we consider five
space-time dimensions, it also describes, like the 5-dimension Mawell field,
3 on-shell degrees of freedom and it settles down an interesting dual
electrodynamics. In view of these aspects, the analysis of both the KR and
Mawell fields in $D$ dimensions, with $d$ ($\leq D$) compactified
dimensions, may be of relevance and motivates us to pursue further
investigations.

\textbf{Ackowledgements}

The Authors thank CNPq and CAPES (of Brazil) for financial support.

\end{document}